\documentstyle[aps,prl,epsfig]{revtex}

\begin{document}
\twocolumn[\hsize\textwidth\columnwidth\hsize\csname @twocolumnfalse\endcsname

\title{Statistics of resonances and delay times: A criterion for Metal-Insulator 
transitions}
\author{Tsampikos Kottos$^{1}$, and Matthias Weiss$^{2}$
\\
$^{1}$Max-Planck-Institut f\"ur Str\"omungsforschung, Bunsenstra\ss e 10,
D-37073 G\"ottingen, Germany\\
$^{2}$ EMBL, Meyerhofstr.~1, D-69117 Heidelberg, Germany
}
\maketitle

\begin{abstract}
We study the distributions of the normalized resonance widths ${\cal P} 
({\tilde \Gamma})$ and delay times ${\cal P} ({\tilde \tau})$ for $3$D disordered 
tight-binding systems at the metal-insulator transition (MIT) by attaching leads 
to the boundary sites. Both distributions are scale invariant, independent of the 
microscopic details of the random potential, and the number of channels. Theoretical 
considerations suggest the existence of a scaling theory for ${\cal P} ({\tilde 
\Gamma })$ in finite samples, and numerical calculations confirm this hypothesis. 
Based on this, we give a new criterion for the determination and analysis of the MIT.
\end{abstract}
\pacs{PACS numbers:03.65.Nk, 72.20.Dp, 71.30.+h, 73.23.-b}
]

Quantum mechanical scattering has been a subject of intensive research activity 
during the last years~\cite{S89,FS97,BGS91,TF00,TC99,SOKG00}, both in Mesoscopic
Physics and in studies of Quantum Chaos. Among the most interesting quantities 
for the description of a scattering process are the Wigner delay times and resonance 
widths. The former quantity captures the time-
dependent aspects of quantum scattering. It can be interpreted as the typical 
time an almost monochromatic wave packet remains in the interaction region. It 
is related to the energy derivative of the total phase shift $\Phi(E)=-i\ln \det 
S(E)$ of the scattering matrix $S (E)$, i.e. $\tau (E) = {1\over M}{d\Phi(E) \over 
dE}$, where $M$ is the number of channels. 
The resonances represent intermediate states with finite lifetimes to which bound 
states of a closed system are converted due to the coupling to a continuum. On 
a formal level, resonances show up as poles of the $S$-matrix occurring at complex 
energies ${\cal E}_n = E_n - \frac i2 \Gamma_n$, where $E_n$ and $\Gamma_n$ are 
called position and width of the resonance, respectively.  

For chaotic/ballistic systems many results are known concerning the distributions
of resonance widths ${\cal P}(\Gamma)$ and Wigner delay times ${\cal P}(\tau)$ 
\cite{FS97}. Recently, the interest has extended to systems showing diffusion and 
localization, where ${\cal P} (\Gamma)$ \cite{BGS91,TF00} and ${\cal P} (\tau)$ 
\cite{TC99} were found to follow universal distributions with algebraic decay. 
At the same time, an attempt to understand systems at critical conditions, was done 
in \cite{SOKG00} where ${\cal P} (\Gamma)$ and ${\cal P} (\tau)$ for one-dimensional 
(1D) quasi-periodic systems was studied. 

Despite the progress in understanding the scattering process of various systems, a 
significant class was left out of the investigation. These are systems, whose closed 
analogues show a metal-insulator transition (MIT) as an external parameter changes. 
In the metallic regime, the eigenstates are extended and the statistical properties 
of the spectrum are quite well described by random matrix theory~\cite{SSSLS93}. 
In particular, the level spacing distribution is very well fitted by the Wigner 
surmise. Deep in the localized regime, the levels become uncorrelated leading to 
a Poissonian level spacing distribution, and the eigenfunctions are exponentially 
localized. At the MIT the eigenfunctions exhibit multifractal behavior with strong 
fluctuations on all scales while the eigenvalue statistics is characterized by a 
third universal distribution \cite{SSSLS93,AS86}. At the same time a considerable
effort was made to understand the shape of the conductance distribution
${\cal P}(g)$ at MIT \cite{A58,RMS01,S90}. However, it is still unclear whether the 
limiting ${\cal P}(g)$ is entirely universal, i.e. dependent only on the 
dimensionality and symmetry class, as required by the one-parameter scaling 
theory of localization \cite{A58}. The latter is one of the major achievements 
in the long history of studying the MIT. Its basic assumption is that close to 
the MIT the change of the conductance $g$ with the sample size $L$ depends only 
on the conductance itself, and not separately on energy, disorder, size and 
shape of the sample, the mean free path etc.

The most prominent realization of systems which undergo a MIT (with 
increasing strength of disorder) is the three-dimensional (3D) Anderson model 
\cite{A58}. Although a lot of studies have been devoted to the analysis of 
eigenfunctions and eigenvalues \cite{SSSLS93,AS86,A58} and of conductance 
\cite{RMS01,S90} at the MIT, the properties of delay times and resonances of 
the $S-$matrix were left unexplored.

Here, for the first time we address this issue and present consequences of the MIT 
on the statistical properties of the (properly) rescaled resonance widths ${\tilde 
\Gamma}$ and delay times ${\tilde \tau}$. We find that they follow a new {\it universal}
distribution, i.e. independent of the microscopic details of the random potential,
and number of channels $M$~\cite{note}. Specifically, they decay asymptotically with powers 
which are different from those found for diffusive or localized systems 
\cite{BGS91,TF00,TC99}, i.e.
\begin{eqnarray}
\label{MIT}
{\cal P}({\tilde \Gamma})&\sim &{\tilde \Gamma}^{-(1+1/3)}\,\,\,;\,\, {\tilde
\Gamma}
\equiv \Gamma/\Delta \nonumber\\
{\cal P}({\tilde \tau})  &\sim &{\tilde \tau}^{-2.5}\,\,\,\,\,\,\,\,\,\,\,\,\,
\,;\, {\tilde \tau} \equiv \tau M \Delta
\end{eqnarray}
where $\Delta$ is the mean level spacing. We relate the power-law decay of ${\cal P} 
({\tilde \Gamma})$ to the anomalous diffusion at the MIT. Finally, based on the 
resonance width distribution, we suggest a new method for determining and analyzing 
the emergence of the MIT and propose a scaling theory near the critical point. Our 
results are confirmed by numerical calculations and are supported by theoretical arguments. 

We consider a 3D sample of volume $L^3$. To each site of the layer $n_x=1$ we attach 
$M= L^2$ semi-infinite single-mode leads, the simplest possible multichannel scattering 
set up~\cite{note2}. The mathematical model that describes the sample is the tight-binding 
Hamiltonian (TBH)
\begin{equation}
\label {tbh}
H_L=\sum_{\bf n} |{\bf n}\rangle V_{\bf n}\langle {\bf n}| + \sum_{{\bf (n,m)}}
|{\bf n}\rangle \langle {\bf m}|
\end{equation}
where ${\bf n}\equiv (n_x,n_y,n_z)$ labels all the $N=L^3$ sites of a cubic lattice,
while the second sum is taken over all nearest-neighbor pairs ${\bf (n,m)}$ on the
lattice. The on-site potential $V_{\bf n}$ for $1\leq n_x,n_y,n_z\leq L$ is
independently and identically distributed with probability ${\cal P}(V_{\bf n})$.
We study three different distributions for the random potential: (a) A box 
distribution, i.e. the $V_{\bf n}$ are uniformly distributed on the interval $[-V/2,
V/2]$, (b) a Gaussian distribution with zero mean and variance $V^2/12$, and (c)
a Cauchy distribution ${\cal P}(V_{\bf n})={V\over \pi (V_{\bf n}^2+V^2)}$. 
For the system defined by Eq.~(\ref{tbh}) the MIT for $E\simeq 0$ occurs for $V=V_c$ 
with (a) $V_c\simeq 16.5$, (b) $V_c\simeq 21.3$, and (c) $V_c\simeq 4.26$ \cite{RMS01}. 
Each lead is described by a 1D semi-infinite TBH
\begin{equation}
\label{leads}
H_M=\sum_{n=1}^{-\infty} (|n><n+1| + |n+1><n|)\,\,.
\end{equation}

Using standard methods one can write the scattering matrix in the form
\cite{D95}
\begin{equation}
\label{smatrix}
S(E) = {\bf 1}-2i \sin (k)\, W^{\,T} \left( E-{\cal H}_{\rm eff}\right)^{-1}W\,\, ,
\end{equation}
where ${\bf 1}$ is the $M\times M$ unit matrix, $k=\arccos(E/2)$ is the wave
vector supported in the leads while ${\cal H}_{\rm eff}$ is an effective
non-hermitian Hamiltonian given by
\begin{equation}
\label{Heff}
{\mathcal{H}}_{\rm eff}=H_L- e^{ik} WW^{\,T}.
\end{equation}
Here $W$ is a $N\times M$ matrix with elements equal to zero or unity. It 
describes at which site of the sample we attach the leads. Moreover, since $\arccos 
(E/2)$ changes only slightly in the center of the band, we put $E=0$ and neglect 
the energy dependence of ${\mathcal{H}}_{\rm eff}$. The poles of $S$ are then 
equal to the complex eigenvalues ${\mathcal E}$ of ${\mathcal{H}}_{\rm eff}$. 

The Wigner delay time $\tau={1\over M}{d\Phi(E)\over dE}$, can be written as
\begin{equation}
\label{tau}
\tau(E) = {1\over M}{\rm Tr} Q(E) \,\,\,;\,\,\,
Q(E)=-iS^{\dagger}(E)\frac{d S(E)} {dE},
\end{equation}
where $Q(E)$ is called Wigner-Smith matrix. Using Eq.~(\ref{smatrix}), we calculated 
the $M\times M$ matrix $Q(E)$ 
\begin{eqnarray}
\label{qmatrix}
Q(E)&=& ({\bf 1}-i \sin (k)\, K)^{-1} \left( {\rm cot}(k) K + 2 \sin (k) W^{\dagger}
{\cal G}\Sigma {\cal G} W \right)\nonumber\\
&&({\bf 1}+i\sin (k)\, K )^{-1}\,\,\,,
\end{eqnarray}
where $K=W^{\dagger}{\cal G} W$, ${\cal G}=(E\Sigma-H_L)^{-1}$ and $\Sigma = {\bf 1}
+{1\over 2} WW^{\dagger}$.  Relation~(\ref{qmatrix}) is very convenient for numerical 
calculations since it anticipates the numerical differentiation which is a rather unstable 
operation. The trace of $Q$ is associated with the density of states of the closed system~\cite{S89} 
and the mean Wigner delay is given by
\begin{equation}
\label{tmean}
<\tau> = {2\pi\over M \Delta}\,\,.
\end{equation}
In order to investigate the distributions of resonance widths ${\tilde \Gamma}$
and delay times ${\tilde \tau}$, we consider the integrated distribution
\begin{equation}
\label{int1}
{\cal P}_{\rm int}(x)= \int_{x}^{\infty} {\cal P}(x')dx' \,\,,
\end{equation}
whose derivative ${\cal P}(x) = -d{\cal P}_{\rm int}/dx$ determines the probability
density of resonance widths ${\cal P}(x={\tilde \Gamma})$ and delay times
${\cal P}(x={\tilde \tau})$, respectively. The maximum size of the matrices that we 
used in our analysis was $N=8000$ ($L=20$). For better statistics a considerable 
number of different disorder realizations has been used. In all cases we had at 
least $10000$ data for statistical processing.
\begin{figure}
\begin{center}
    \epsfxsize=8.4cm
    \leavevmode
    \epsffile{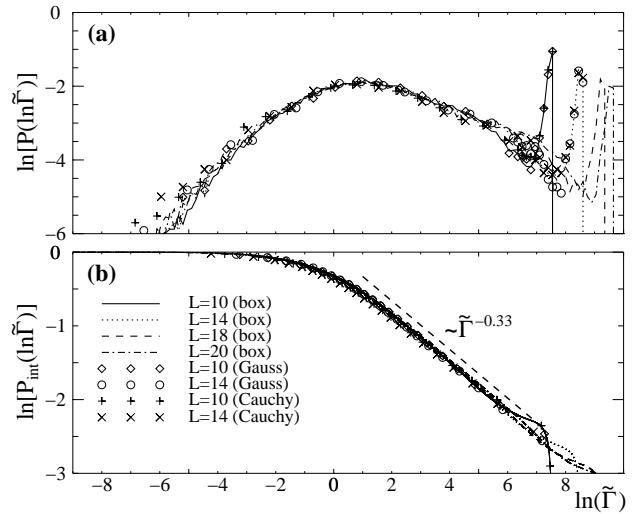}
\caption{(a) Universal behavior of ${\cal P}({\tilde \Gamma})$ at the MIT (reported 
here as ${\cal P}(\ln({\tilde \Gamma}))$) for various sample sizes $L$ and potential 
distributions. (b) The integrated distribution ${\cal P}_{\rm int}({\tilde \Gamma})$ 
asymptotically decays as ${\cal P}_{\rm int}({\tilde \Gamma}) \sim {\tilde \Gamma}^{-
0.333}$ (dashed line).}
\label{fig:fig2}
\end{center}
\end{figure}
Figure~\ref{fig:fig2}a displays our results for the distribution of the logarithm 
of the rescaled resonance widths ${\cal P}(\ln({\tilde \Gamma}))$ \cite{loga} for 
the three different distributions ${\cal P}(V_{\bf n})$ of the random potential and 
for various sample sizes $L$. The body of the distribution 
function in all cases coincides and does not change its shape or width. Of course, 
the far tail of this universal distribution develops better with increasing $L$. 
The sharp peak appearing at the right is an artifact of our 
choice to neglect the energy dependence of ${\cal H}_{\rm eff}$. Taking the energy 
dependence into account results in a much milder behavior (see for example Figs.~4,6
in ref. \cite{TF00}). We conclude therefore that at the MIT, the distribution of 
rescaled resonances is indeed scale-invariant independent of the microscopic details 
of the potential.

Next we turn to the quantitative analysis of ${\cal P}({\tilde \Gamma})$. To this end 
we analyze numerically the integrated distribution ${\cal P}_{\rm int}({\tilde \Gamma})
$. Our results are reported in Fig.~\ref{fig:fig2}b. An inverse power law ${\cal P}_{
\rm int}({\tilde \Gamma}) \sim {\tilde \Gamma}^{ -\alpha}$ is evident. The best fit to 
the numerical data yields $\alpha=0.333\pm 0.005$ in accordance with Eq.~(\ref{MIT}). 
The behavior of the statistically insignificant extreme large $\Gamma$ tails of
${\cal P}({\tilde \Gamma})$ (associated with the sharp peak on the right) is 
essentially determined by the coupling to the leads, and therefore is model-dependent. 
Nevertheless, it is reasonable to assume that the relative number of resonances 
contributing to the peak is proportional to $M/L^3 \sim L^{-1}$ and thus these 
extreme non-universal far tails subside as $L$ increases in agreement with our
data (see Fig.~\ref{fig:fig2})
\begin{figure}
\begin{center}
    \epsfxsize=8.4cm
    \leavevmode
    \epsffile{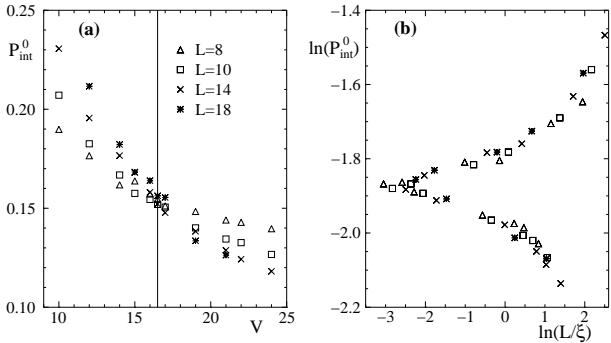}
\caption{(a) ${\cal P}_{\rm int}^0(V,L)$ as a function of $V$ for different 
system sizes $L$ provides a means to determine the critical point $V_c$ of the MIT 
(vertical line at $V=16.5$). (b) The one-parameter scaling of ${\cal P}_{\rm int}^0(V,L)$ 
[Eq.~(\ref{scale})] is confirmed for various system sizes $L$ and disorder strengths $V$
using the box distribution. 
}
\label{fig:fig3}
\end{center}
\end{figure}
The following heuristic argument provides some understanding of the universal 
character of ${\cal P}(\Gamma)$ and gives a quantitative estimate for
the power law (\ref{MIT}). At this point we wish to generalize our analysis to 
$d$-dimensional systems with a MIT. The specific case of $d=3$ will be recovered at 
the end. At $V=V_c$ the conductance of a $d$-dimensional sample has a finite value 
$g_{\rm c}\sim 1$. Approaching the MIT from the metallic side one has $g\sim E_T/ 
\Delta$, where $E_T=D/R^2$ is the Thouless energy, $D$ is the diffusion coefficient, 
and $\Delta \sim 1/R^d$ is the mean level spacing in a $d$-dimensional sample with 
linear size $R$. This yields $D\sim g_{\rm c}/R^{d-2}$ at $V_c$. Taking into
account that $D=R^2/t$, we get for the spreading of an excitation at the MIT
\begin{equation}
\label{trpert}
R^d(t)\sim g_{\rm c} t\,\,.
\end{equation}
We now consider the influence of the leads on the spectrum of an isolated sample
(for simplicity we consider that the leads are attached to the sites of the layer 
$n_x=1$, but the argument can be generalized easily). Due to the leads the levels 
are broadened by a width $\Gamma$ and the eigenstates are turned into resonant states. 
The inverse of $\Gamma$ represents the quantum lifetime of an electron in the 
corresponding resonant state before it escapes out into the leads. We 
suppose that the resonant states are uniformly distributed along the sample. 
Their lifetime can be associated with the typical time $t_R\sim 1/ \Gamma_R\sim R^d/g_{\rm 
c}$ [Eq.~(\ref{trpert})] a particle needs to reach the boundary $n_x=1$, when starting 
at distance $R$ apart. Moreover we assume that the particles are uniformly distributed
inside the sample. This classical picture can be justified for all states
with $\Gamma_R > \Gamma_L=1/t_L = L^d/g_{\rm c}\sim \Delta$ since the level 
discreteness is unimportant. Therefore the number of states 
with $\Gamma>\Gamma_R$ is 
\begin{equation}
\label{igam}
{\cal P}_{\rm int}(\Gamma_R)\approx R L^{d-1}/L^d=R/L \sim 
(g_{\rm c}/\Gamma_R)^{1/d}/L\,\,.
\end{equation}
By repeating the same argument for $R\to L$ (i.e. $\Gamma_R\to \Gamma_L$) we finally 
obtain for all $\Gamma\ge\Gamma_L$
\begin{equation}
\label{final}
{\cal P}_{\rm int}(\Gamma) \sim g_{\rm c}^{1/d} (\Delta/\Gamma)^{1/d}
\Leftrightarrow
{\cal P}({\tilde \Gamma})\sim g_{\rm c}^{1/d}  {\tilde \Gamma}^{-(1+1/d)}\,\, ,
\end{equation}
which is equivalent to Eq.~(\ref{MIT}) for the case $d=3$. Since $g_{\rm 
c}$ is universal at the MIT \cite{RMS01}, we expect ${\cal P}({\tilde \Gamma})$
to be universal as well. An additional conclusion is that the power-law decay of 
${\cal P}({\tilde \Gamma})$ is determined by the anomalous diffusion at the MIT, 
which leads to the dependence on the dimensionality of the system.

In the original proposal of the scaling theory of localization, the conductance $g$ 
is the relevant parameter \cite{A58}. A manifestation of this statement is seen in 
Eq.~(\ref{final}) where ${\cal P}_{\rm int}^{0}\equiv {\cal P}_{\rm int}({\tilde \Gamma}_0
)$ is proportional to the conductance $g$. It is therefore natural to expect that 
${\cal P}_{\rm int}^{0}$ will follow a scaling behavior for finite $L$ (and for some 
${\tilde \Gamma}_0\sim 1$), that is similar to the one obeyed by the conductance $g$. We
therefore postulate the following scaling hypothesis
\begin{equation}
\label{scale}
{\cal P}_{\rm int}^{0}(V,L)=f(L/\xi(V))\,\, ,
\end{equation}
where $\xi(V)$ is the correlation length of the MIT. In the insulating phase ($V>
V_c$) the conductance of a sample with length $L$ behaves as $g(L)\sim \exp(-L/\xi)$ 
due to the exponential localization of the eigenstates, and therefore we have $g(L_1) 
< g(L_2)$ for $L_1>L_2$. Based on Eq.~(\ref{final}) we expect the same behavior 
for ${\cal P}_{\rm int}^{0}$ i.e. for every finite $L_1>L_2$ we must have ${\cal P}_{\rm 
int}^{0}(V,L_1)<{\cal P}_{\rm int}^{0}(V,L_2)$. On the other hand, in the metallic 
regime ($V<V_c$) we have that $g(L)=D L$ and therefore we expect from Eq.~(\ref{final}) 
${\cal P}_{\rm int}^{0}(V,L_1)>{\cal P}_{\rm int}^{0}(V,L_2)$. Thus, the critical 
point is the one at which the size effect changes its sign, or in other words, 
the point where all curves ${\cal P}_{\rm int}^{0}(V,L)$ for various $L$ cross.
These considerations are confirmed in Fig.~\ref{fig:fig3}a, where we report 
as an example the evolution of ${\cal P}_{\rm int}^{0}(V)$ for different $L$ using
the box distribution. From our data we determine the critical point to be 
$V=V_c=16.5\pm 0.5$ in agreement with other calculations \cite{A58,RMS01}.
In order to verify the scaling hypothesis (\ref{scale}) we report in Fig.~\ref{fig:fig3}b 
the same data as a function of the scaling ratio $L/\xi$. All points collapse on two separate 
branches for $V<V_c$ and $V>V_c$. We consider this to be a good confirmation of the scaling 
hypothesis.

We now turn to the analysis of the delay time statistics ${\cal P}({\tilde \tau })
$. Equation~(\ref{tmean}) provides us with a simple understanding of the scaling
${\tilde \tau}=\tau \Delta M =\tau/<\tau>$ proposed in (\ref{MIT}). In Fig.~4a we 
report the distribution of the logarithm of the rescaled delay times ${\cal P}
(\ln({\tilde \tau}))$ \cite{loga} for the three different distributions ${\cal P}
(V_{\bf n})$ of the random potential and for various sample sizes $L$. In all cases 
the distribution functions lie one on top of each other, indicating that their 
shape is universal. Together with our previous result for ${\cal P}(\Gamma)$ this 
indicates that at the MIT only one fundamental energy scale, the mean level spacing 
$\Delta$, is required to characterize the statistical properties of the $S$-matrix.

\begin{figure}
\begin{center}
    \epsfxsize=8.4cm
    \leavevmode
    \epsffile{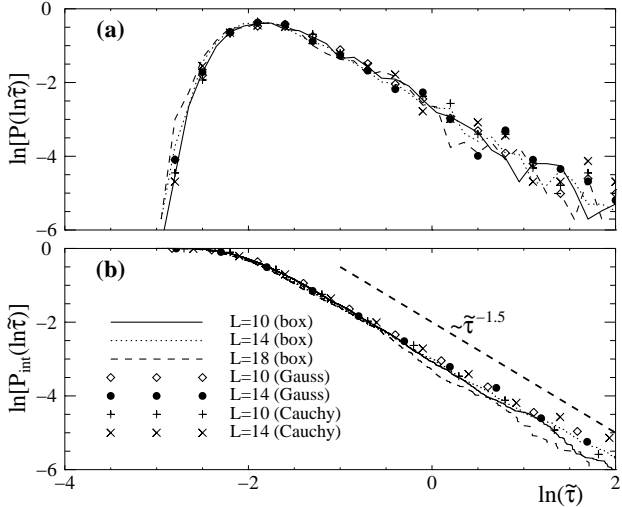}
\caption{(a) Universal behavior of ${\cal P}({\tilde \tau})$ at the MIT (reported 
here as ${\cal P}(\ln({\tilde\tau}))$) for various sample sizes $L$ and potential 
distributions. (b) The integrated distribution ${\cal P}_{\rm int}({\tilde \tau})$ 
asymptotically decays as ${\cal P}_{\rm int}({\tilde \tau}) \sim {\tilde \tau}^{-
1.5}$ (dashed line).
}
\label{fig:fig4}
\end{center}
\end{figure}

We now discuss the general features of ${\cal P}({\tilde \tau})$. For $\tilde\tau\ll 
1$ the behavior of ${\cal P} ({\tilde \tau})$ can be deduced from Fig.~\ref{fig:fig4}a, 
which clearly shows that $\ln{\cal P} (\ln{\tilde \tau})$ decreases faster than $\ln{
\tilde \tau}$. This guarantees that ${\cal P}({\tilde \tau}\rightarrow 0) \rightarrow 
0$ since ${\cal P} (\tilde \tau\to0)=c\neq0$ implies $\ln{\cal P}(\ln{\tilde \tau})=
\ln(c) + \ln({\tilde \tau})$. Thus, a gap is formed at the origin of ${\cal P} ({\tilde \tau})$. 
We point out that a similar behavior is predicted for the conductance distribution at 
the MIT by the $\epsilon$-expansion in field theory \cite{S90} and has been verified 
by recent numerical calculations \cite{RMS01}.

In Fig.~\ref{fig:fig4}b we present ${\cal P}_{\rm int}({\tilde \tau})$ in a double 
logarithmic plot. Again all curves coincide. Moreover the tails show 
a power-law decay ${\cal P}_{\rm int}({\tilde \tau}) \sim {\tilde \tau}^{-\gamma}$ 
with $\gamma\approx 1.5\pm 0.05$ given by a least square fit. The fact that $\gamma
> 2$ indicates that a mean delay time can be defined at the MIT in contrast to the 
localized regime where we have divergence of the first moment of ${\cal P}(\tau)$
\cite{TC99}. This is yet another manifestation of the fact that at MIT 
the conductance has a 
well defined value $g_{\rm c}$. We note the divergence of higher moments of ${\cal P}
({\tilde \tau})$ and point out that a similar behavior is predicted for the higher 
cumulants of the conductance distribution at the MIT from field theoretical 
calculations~\cite{S90}. 

We thank Y. Fyodorov and L.~Hufnagel for useful comments. (T.K) thanks U. Smilansky 
for initiating his interest in quantum scattering. (M.W) acknowledges financial 
support by an EMBO fellowship.


\begin{thebibliography}{99}
\bibitem{loga} We plot ${\cal P}(\ln(x))$,
$x={\tilde \Gamma},{\tilde \tau}$, instead of ${\cal P}(x)$ because
we want to stress the universality of the distributions in the whole range from
$x\ll 1$ to $x\gg 1$.
\end{thebibliography}
\end{document}